\documentclass[twocolumn,secnumarabic,amsmath, amssymb, nobibnotes, aps, prl]{revtex4-1}
\usepackage{listings}

\usepackage{subfigure}
\usepackage{graphicx}
\graphicspath{{./}{./images/}{./images_all/}{./images_all/spin_pumping_circuit/}}

\usepackage{dcolumn}
\usepackage{bm}

\begin{document}

\title{Spin circuit representation of spin pumping}

\author{Kuntal Roy}
\email{royk@purdue.edu.}
\affiliation{School of Electrical and Computer Engineering, Purdue University, West Lafayette, Indiana 47907, USA}


\begin{abstract}
Circuit theory has been tremendously successful in translating physical equations into circuit elements in organized form for further analysis and proposing creative designs for applications. With the advent of new materials and phenomena in the field of spintronics and nanomagnetics, it is imperative to construct the spin circuit representations for different materials and phenomena. Spin pumping is a phenomenon by which a pure spin current can be injected into the adjacent layers. If the adjacent layer is a material with a high spin orbit coupling, a considerable amount of charge voltage can be generated via inverse spin Hall effect, allowing spin detection. Here we develop the spin circuit representation of spin pumping. We then combine it with the spin circuit representation for the materials having spin Hall effect to show that it reproduces the standard results in literature. We further show how complex multilayers can be analyzed by simply writing a netlist.
\end{abstract}


\maketitle

Spin pumping~\cite{RefWorks:1311,RefWorks:1305,*RefWorks:1306,*RefWorks:1304,*RefWorks:1307,*RefWorks:1308,*RefWorks:1309,*RefWorks:1310,RefWorks:881,*RefWorks:1041,RefWorks:876,RefWorks:882,RefWorks:989}, unlike charge pumping~\cite{RefWorks:1034,*RefWorks:1324}, injects a \emph{pure} spin current into surrounding conductors. If the adjacent material possesses high spin-orbit coupling~\cite{RefWorks:757,RefWorks:818,RefWorks:814,*RefWorks:813,RefWorks:817,*RefWorks:608,*RefWorks:758,*RefWorks:1341,*RefWorks:755,*RefWorks:1342,*RefWorks:1343}), a considerable amount of charge voltage can be generated allowing the detection of spin current via inverse spin Hall effect (ISHE)~\cite{RefWorks:1325,RefWorks:1111,RefWorks:1101,RefWorks:1278}. According to Onsager's reciprocity~\cite{RefWorks:1292,*RefWorks:1293,RefWorks:1071}, spin pumping and ISHE are the reciprocal phenomena~\cite{RefWorks:1295} of spin momentum transfer~\cite{RefWorks:8,*RefWorks:155,*RefWorks:7,*RefWorks:196} and direct spin Hall effect (SHE)~\cite{RefWorks:760,*RefWorks:764,*RefWorks:1198,RefWorks:771,*RefWorks:765,*RefWorks:902,RefWorks:769,roy14_3}, respectively and it gives us an alternative methodology to understand and estimate the relevant parameters in the system.

Kirchoff's circuit laws (current and voltage laws, referred as KCL and KVL, respectively) have been ubiquitous in the development of the modern transistor-based technology and there are commercial programs for SPICE (Simulation Program with Integrated Circuit Emphasis), e.g., HSPICE~\cite{hspice}. In this way, an equivalent circuit is constructed based on the underlying physical governing equations for simplified understandings and the development of the complex designs~\cite{rabae03}. For spintronic circuits, the voltages and currents at different nodes are of 4-components (1 for charge and 3 for spin vector) and the conductances are $4\times 4$ matrices ($c-z-x-y$ basis). Such representations have been developed earlier e.g., for ferromagnet (FM), normal metal (NM), FM-NM interface, spin Hall effect (SHE)~\cite{RefWorks:198,srini14,RefWorks:1253}. 

Here, we construct the spin circuit representation of spin pumping. We deduce the 3-component version since spin pumping injects \emph{pure} spin current (without any charge component) and show that it can reproduce the established expressions in literature for effective spin mixing conductance considering diffusive NM~\cite{RefWorks:876} and the inverse spin Hall voltage due to spin pumping~\cite{RefWorks:814,*RefWorks:813}. We further employ such spin circuit for a complex structure of multilayers and show that it simply reduces to an equivalent circuit matching the mathematically derived expression in literature~\cite{RefWorks:1090}. Furthermore, we show how to write a simple netlist to solve and derive the effective spin mixing conductance for complex structures.  

The expressions of spin current due to spin pumping~\cite{RefWorks:881,*RefWorks:1041,RefWorks:876} and spin battery~\cite{RefWorks:878,RefWorks:198} due to a precessing magnetization $\hat{m}$ can be written, respectively as
\begin{equation}
\vec{I}_{SP} = \cfrac{\hbar}{2e}\, \left(2G_r^{\uparrow \downarrow}\, \hat{m} \times \frac{d\hat{m}}{dt} + 2G_i^{\uparrow \downarrow}\, \frac{d\hat{m}}{dt} \right)
\label{eq:vector_Isp}
\end{equation}
and
\begin{equation}
\vec{V}_{SP} = \frac{\hbar}{2e} \left(\hat{m} \times \frac{d\hat{m}}{dt}\right),
\label{eq:vector_Vsp}
\end{equation}
where $G^{\uparrow \downarrow}$ (= $G_r^{\uparrow \downarrow} + i\,G_i^{\uparrow \downarrow}$) is the complex (reflection) spin mixing conductance at the ferromagnet-normal metal (FM-NM) interface~\cite{RefWorks:1039,*RefWorks:1040,RefWorks:1018}.

\begin{figure*}
\centering
\includegraphics{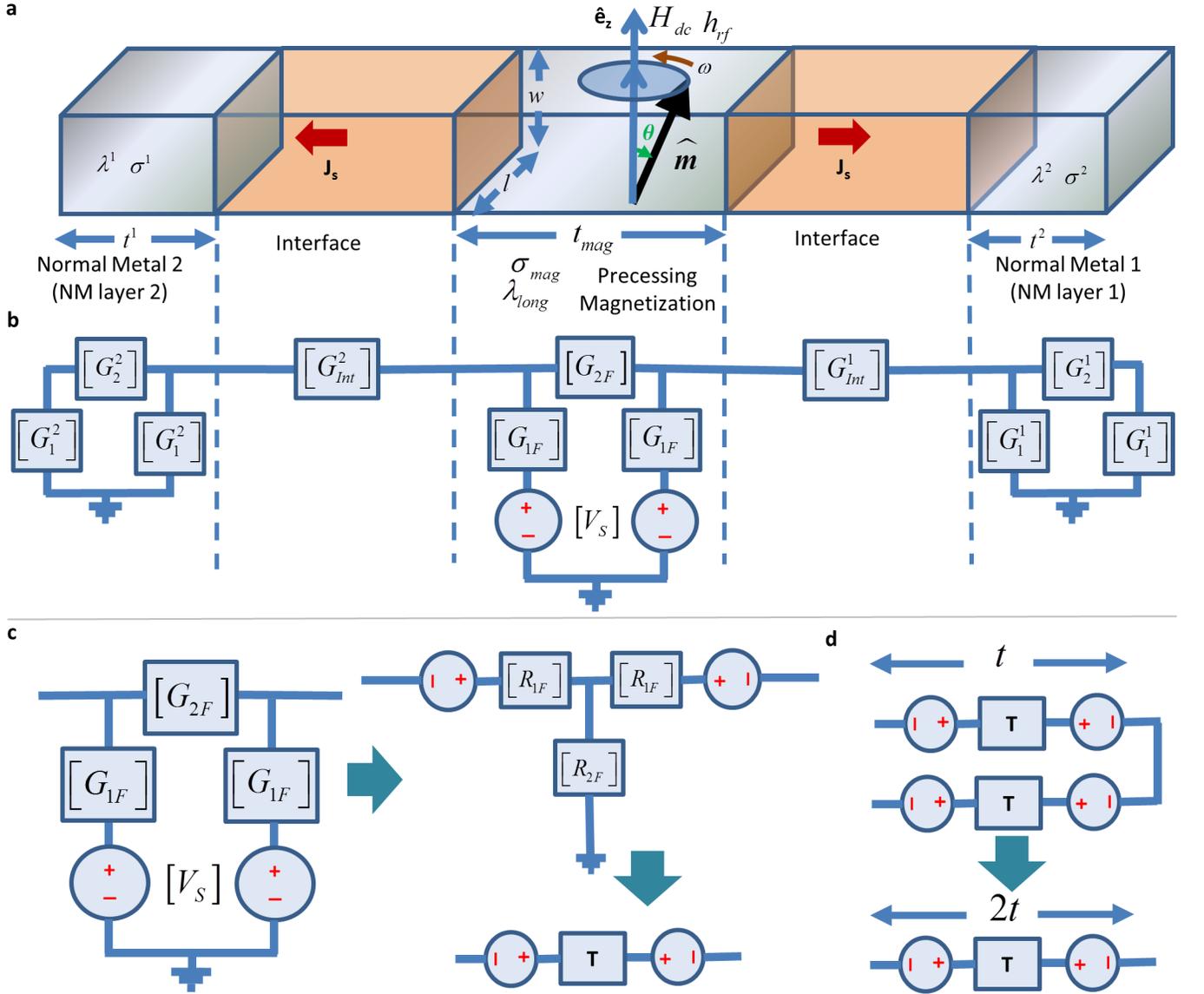}
\caption{\label{fig:spin_pumping} (a) A precessing magnetization in a magnetic layer with uniform mode of excitation. This can generate a pure spin current to the adjacent normal metals (NMs). $J_s$ is the spin current density, $H_{dc}$ is the applied dc magnetic field, $h_{rf}$ is the rf driving field, $\omega$ and $\theta$ are the precession frequency and angle, respectively. (b) Spin circuit representation of spin pumping for ferromagnet (FM), NM, and FM-NM interface. The voltage source $[V_{S}]$ acts as a spin battery, $[G_{Int}^{1(2)}]$ is the FM-NM interfacial spin mixing conductances for the two interfaces. (c) The FM $\pi$-circuit can be converted to an equivalent $T$-circuit. (d) Two $T$-circuits can be joined to get an equivalent $T$-circuit of twice length. Note that the voltage sources in between the two $T$-circuits get canceled out and thus the spin battery only appears at the interfaces.}
\end{figure*}

\begin{figure*}
\centering
\includegraphics{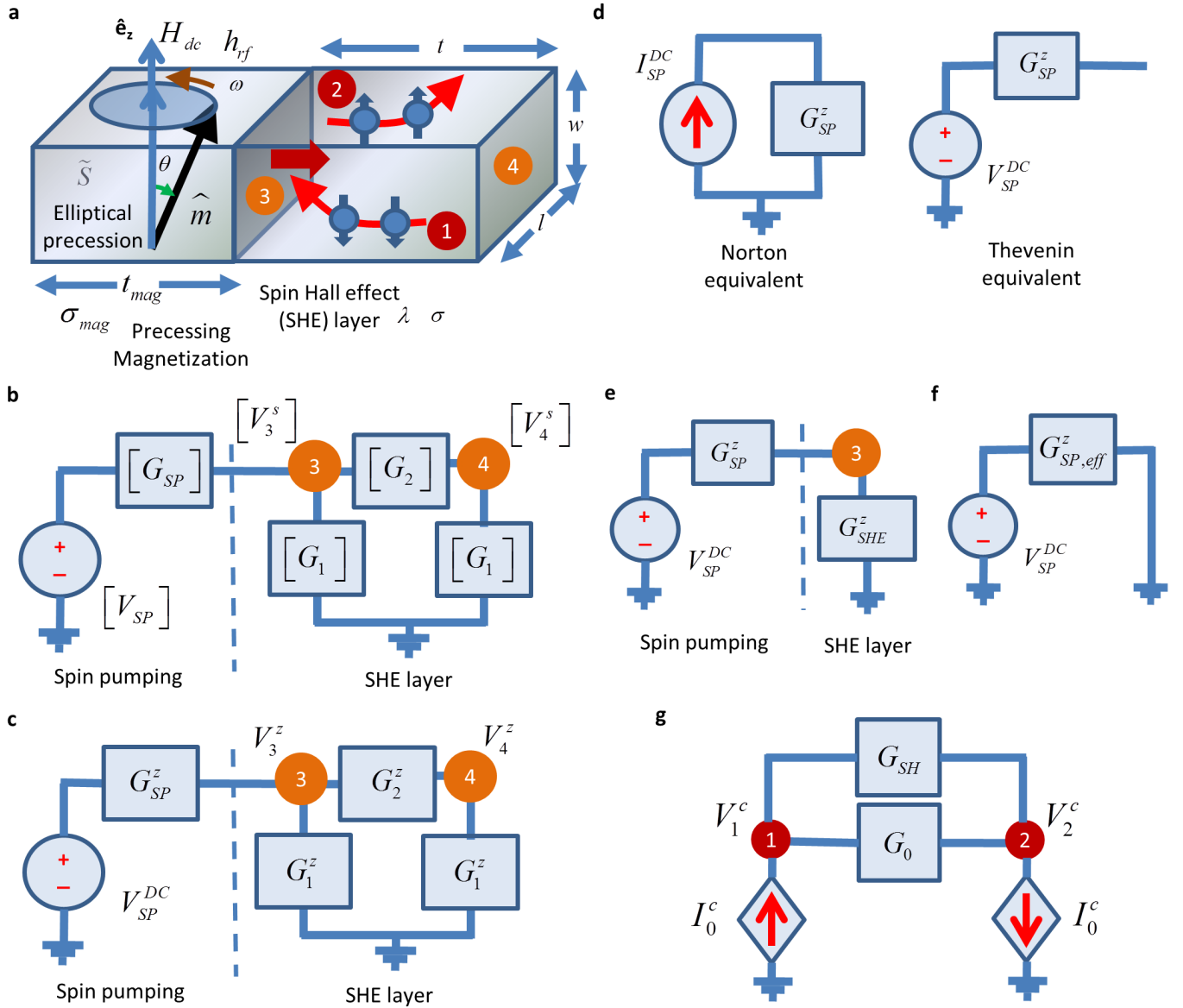}
\caption{\label{fig:spin_pumping_benchmark} (a) A precessing magnetization is pumping pure spin current to an adjacent layer possessing a high spin orbit coupling and it generates a considerable amount of charge current due to inverse spin Hall effect (ISHE). Charge potentials are developed at the surfaces marked by 1 and 2, while spin potentials are developed at the surfaces marked by 3 and 4. (b) Instantaneous 3-component spin circuit with the voltage source $[V_{SP}]$ acts as a spin battery, $[G_{SP}]$ is the interfacial spin mixing conductance between the magnetic layer and the SHE layer. (c) A dc spin circuit with average spin polarization acting in the $z$-direction. (d) The spin circuit for the spin pumping can be represented by either Thevenin-equivalent with a spin battery or Norton-equivalent with a spin current source. (e) Reduced spin circuit with the SHE layer conductance represented by a single conductance. (f) Spin circuit representing the effective spin mixing conductance with the SHE layer conductance included in it. (g) The charge circuit for the spin-to-charge conversion by ISHE with the current sources $I_0^c$ dependent on the spin circuit in the part (c), $G_0$ and $G_{SH}$ are the conductances for the SHE and FM layers, respectively.}
\end{figure*}

\begin{figure*}
\centering
\includegraphics{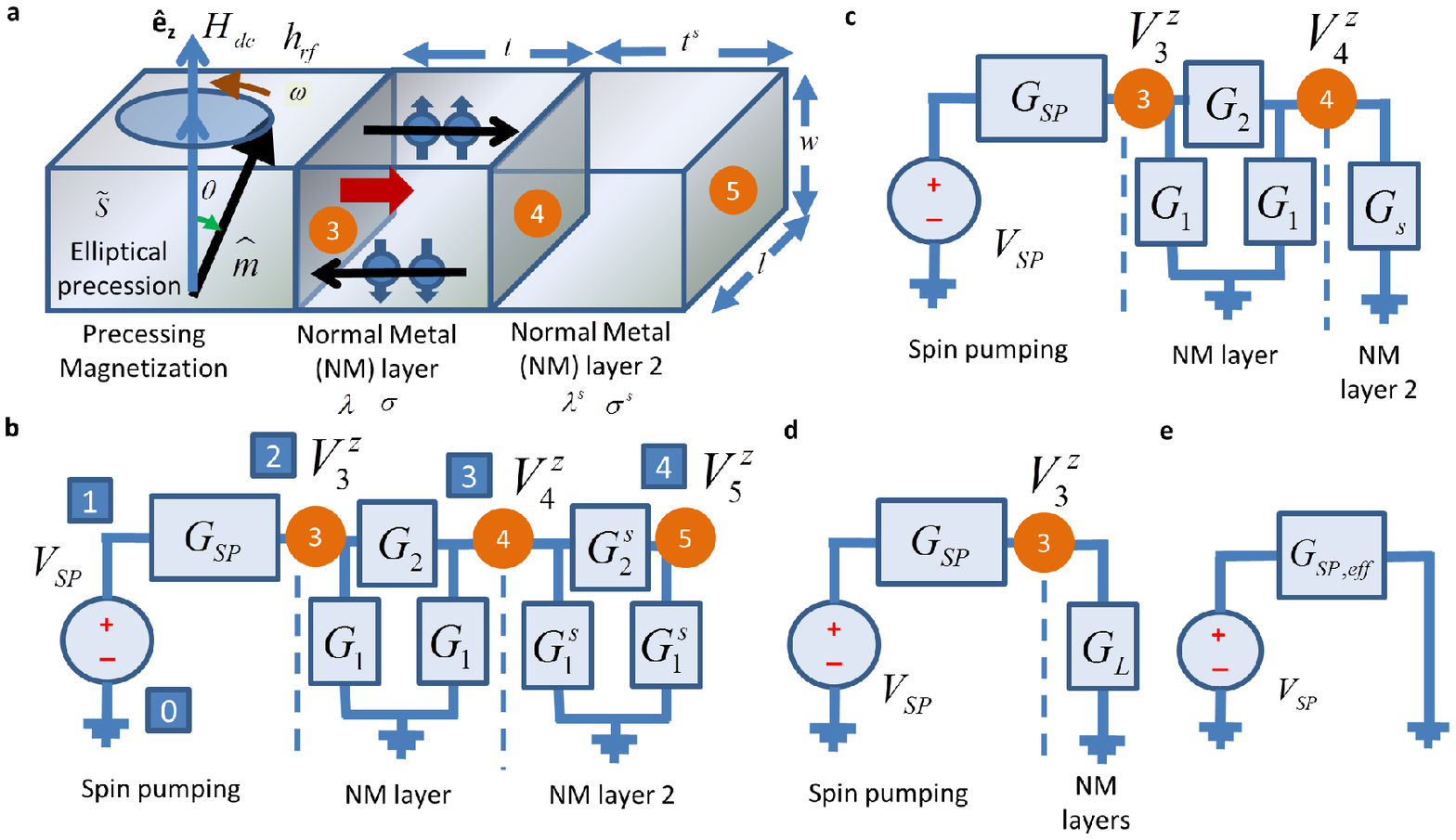}
\caption{\label{fig:spin_pumping_multilayer} (a) A precessing magnetization is pumping spins into two subsequent layers. (b) Spin circuit representation of the corresponding layers. (c) Reduced spin circuit representation with a single conductance $G_s$ representing the second layer. (d) Reduced spin circuit with a single conductance $G_L$ representing both the layers. (e) Spin circuit representing the effective spin mixing conductance with including the spin conductances of the multilayers.}
\end{figure*}

Figure~\ref{fig:spin_pumping} shows the spin circuit representation of spin pumping, with the conductances~\cite{RefWorks:1284,*RefWorks:1289,*RefWorks:1290} (after the necessary modifications~\cite{RefWorks:198,srini14}) and spin voltage sources represented as follows.
\begin{equation}
\left\lbrack G_{1}^{1(2)} \right\rbrack= \left( \begin{array}{cccc}
  0 & 0 & 0 & 0\\ 
  0 & G_1^{1(2)} & 0 & 0\\
	0 & 0 & G_1^{1(2)} & 0\\ 
	0 & 0 & 0 & G_1^{1(2)}
\end{array} \right),
\label{eq:G1}
\end{equation}
\begin{equation}
\left\lbrack G_{2}^{1(2)} \right\rbrack= \left( \begin{array}{cccc}
  G_{n}^{1(2)} & 0 & 0 & 0\\ 
  0 & G_2^{1(2)} & 0 & 0\\
	0 & 0 & G_2^{1(2)} & 0\\ 
	0 & 0 & 0 & G_2^{1(2)}
\end{array} \right),
\label{eq:G2}
\end{equation}
\begin{equation}
\left\lbrack G_{1F} \right\rbrack= \left( \begin{array}{cccc}
  0 & 0 & 0 & 0\\ 
  0 & G_{sh}^{z} & 0 & 0\\
	0 & 0 & \infty & \infty\\ 
	0 & 0 & \infty & \infty
\end{array} \right),
\left\lbrack G_{2F} \right\rbrack= \left( \begin{array}{cccc}
  G_{f} & P G_f & 0 & 0\\
  P G_f & G_{se}^{z} & 0 & 0\\
	0 & 0 & 0 & 0\\ 
	0 & 0 & 0 & 0
\end{array} \right),
\label{eq:G_1F_2F}
\end{equation}
\begin{equation}
\left\lbrack G_{Int}^{1(2)} \right\rbrack= \left( \begin{array}{cccc}
  G_I^{1(2)} & P_I^{1(2)} G_I^{1(2)} & 0 & 0\\ 
  P_I^{1(2)} G_I^{1(2)} & G_I^{1(2)} & 0 & 0\\
	0 & 0 & 2G_r^{\uparrow \downarrow \, 1(2)} & 2G_i^{\uparrow \downarrow \, 1(2)}\\ 
	0 & 0 & 2G_i^{\uparrow \downarrow \, 1(2)} & 2G_r^{\uparrow \downarrow \, 1(2)}
\end{array} \right),
\label{eq:GInt}
\end{equation}
and $\left\lbrack V_S \right\rbrack = \left\lbrack 0 \quad \vec{V}_{SP}\right\rbrack^T$, 
\begin{flushleft}
where $G_{f}=\sigma_{mag} l w/t_{mag}$, $G_{n}^{1(2)}=\sigma^{1(2)} l w/t^{1(2)}$, $G_z=\sigma_z l w/t_{mag}$, $G_{sh}^{z}=G_zQ\,tanh(t_{mag}/2\lambda_{long})$, $G_{se}^{z}=G_z\left(P^2+Q\,csch(t_{mag}/2\lambda_{long})\right)$, $Q=(t_{mag}/\lambda_{long})(1-P^2)$, $G_1^{1(2)}=(\sigma^{1(2)} l w/t^{1(2)}) tanh(t^{1(2)}/2\lambda^{1(2)})$, $G_2^{1(2)}=(\sigma^{1(2)} l w/t^{1(2)}) csch(t/\lambda^{1(2)})$, 
\end{flushleft}
$P$ is the spin polarization of the FM, $G_I$ and $P_I$ are the FM-NM interface conductance and polarization, respectively, $\lambda^{1(2)}$, $\sigma^{1(2)}$, and $t^{1(2)}$ are the spin diffusion length, conductivity, and thickness of the NM$^{1(2)}$ layer, respectively, $\lambda_{long}$, $\sigma_z$, and $t_{mag}$ are the \emph{longitudinal} spin diffusion length, conductivity, and thickness of the FM layer, respectively, and $l \times w$ is the cross-sectional area.

The following points should be noted from the spin circuit representation of spin pumping in the Fig.~\ref{fig:spin_pumping}: (1) We include the spin battery in the FM module since a magnet can be precessed without the connection of an NM module. (2) Figs.~\ref{fig:spin_pumping}(c) and~\ref{fig:spin_pumping}(d) explain (see the caption) that the spin battery appears at the interface only, which accords with the established physical concept in literature~\cite{RefWorks:876}. (3) We consider the transverse components ($x-y$) of the conductances entirely at the FM-NM interface ($[G_{int}^{1(2)}]$), thus the transverse components of $[G_{1F}]$ are $\infty$. (4) We consider that the magnet is thick enough (compared to the \emph{transverse} spin diffusion length $\lambda_{tran}$, which is a few monolayers for the typical transition metals) so that the \emph{transmission} spin mixing conductances are nearly zero, and thus we only consider the \emph{reflection} spin mixing conductances~\cite{RefWorks:876}. (5) For magnetic insulators~\cite{RefWorks:812,*RefWorks:992,*RefWorks:885,*RefWorks:1189,*RefWorks:991,*RefWorks:1017,*RefWorks:998}, $\sigma_{mag}=0$, $P=0$, $\sigma_z$ ($\lambda_{long}$) represents the spin wave/magnon conductivity (diffusion length), and the \emph{transmission} spin mixing conductances are \emph{exactly} zero (i.e., the transverse components of $[G_{1F}]$ and $[G_{2F}]$ are \emph{exactly} $\infty$ and $0$, respectively).

Figure~\ref{fig:spin_pumping_benchmark} shows the 3-component version of  spin pumping and the generation of inverse spin Hall voltage in the transverse direction due to the symmetry involved in the system~\cite{RefWorks:814,RefWorks:813}. The instantaneous spin pumping [Figure~\ref{fig:spin_pumping_benchmark}(b)] in matrix form can be represented by $\left\lbrack I_{SP} \right\rbrack = \left\lbrack G_{SP} \right\rbrack \left\lbrack V_{SP} \right\rbrack$, where
\begin{equation}
\left\lbrack I_{SP} \right\rbrack= lw \, \tilde{S} \,\left(\cfrac{2e}{\hbar}\right) \,\cfrac{\hbar \omega}{4\pi}\,\left( \begin{array}{c}
 g_{r}^{\uparrow \downarrow} \left(1-m_z^2 \right)\\ 
-g_{r}^{\uparrow \downarrow} m_x m_z - g_{i}^{\uparrow \downarrow} m_y\\
-g_{r}^{\uparrow \downarrow} m_y m_z + g_{i}^{\uparrow \downarrow} m_x
\end{array} \right),
\label{eq:matrix_Isp}
\end{equation}
\begin{equation}
\left\lbrack V_{SP} \right\rbrack=\tilde{S} \,\frac{\hbar \omega}{2e}\,\left( \begin{array}{c}
 \left(1-m_z^2 \right)\\ 
-m_x m_z \\
-m_y m_z 
\end{array} \right),
\label{eq:matrix_Vsp}
\end{equation}
$G_r^{\uparrow \downarrow} = lw\,(e^2/h)\,g_r^{\uparrow \downarrow}$, $G_i^{\uparrow \downarrow} = lw\, (e^2/h)\,g_i^{\uparrow \downarrow}$ ($g^{\uparrow \downarrow} = g_r^{\uparrow \downarrow} + i g_i^{\uparrow \downarrow}$ is the complex spin mixing conductance per unit area), $\tilde{S}$ is the frequency dependent elliptical precession factor due to thin magnetic film~\cite{RefWorks:884}, the components of $\left\lbrack G_{SP} \right\rbrack$  
\begin{align}
	G_{SP}^{\,nn} &= 2G_r^{\uparrow \downarrow} \left(1-m_n^2 \right) \quad n=(x,y,z),  \nonumber\\
	G_{SP}^{xy} (G_{SP}^{yx})&=-2G_{r}^{\uparrow \downarrow} m_x m_z \pm 2G_{i}^{\uparrow \downarrow} m_y, \nonumber\\
	G_{SP}^{yz} (G_{SP}^{zy})&=-2G_{r}^{\uparrow \downarrow} m_x m_y \pm 2G_{i}^{\uparrow \downarrow} m_z, \nonumber\\
	G_{SP}^{zx} (G_{SP}^{xz})&=-2G_{r}^{\uparrow \downarrow} m_y m_z \pm 2G_{i}^{\uparrow \downarrow} m_x, \nonumber
\label{eq:Gsp}
\end{align}
$\left\lbrack G_{1} \right\rbrack=G_\lambda tanh \left(\frac{t}{2\lambda}\right) \left\lbrack I_{3\times3} \right\rbrack$, $\left\lbrack G_{2} \right\rbrack=G_\lambda csch \left(\frac{t}{\lambda}\right) \left\lbrack I_{3\times3} \right\rbrack$, $G_\lambda = \sigma l w/\lambda$, and $\left\lbrack I_{3\times3}\right\rbrack$ is the ${3\times3}$ identity matrix.

The spin circuit representation of average spin pumping for a complete precession is depicted in the Fig.~\ref{fig:spin_pumping_benchmark}(c) with the voltage source (or the current source depicted in the Fig.~\ref{fig:spin_pumping_benchmark}(d)) represented by $V_{SP}^{DC} = \tilde{S}\,\frac{\hbar \omega}{2e} sin^2\theta$ ($I_{SP}^{DC} = lw\,\tilde{S} \, \frac{e \omega}{2\pi}  g_r^{\uparrow \downarrow} sin^2\theta$). Thus $G_{SP}^z = I_{SP}^{DC}/V_{SP}^{DC} = lw (2e^2/h) g_r^{\uparrow \downarrow}$. Note that first principles calculations and experiments have shown that the imaginary component of $g^{\uparrow \downarrow}$ is negligible for metallic interfaces (i.e., $g^{\uparrow \downarrow} \simeq g_r^{\uparrow \downarrow}$)~\cite{RefWorks:1046,RefWorks:814,*RefWorks:813}. $G_1^z = G_\lambda tanh (t/2\lambda)$, $G_2^z = G_\lambda csch (t/\lambda)$. From Fig.~\ref{fig:spin_pumping_benchmark}(d), $G_{SHE}^z = G_1^z + G_1^z G_2^z/(G_1^z + G_2^z) = G_\lambda/coth(t/\lambda)$. From Fig.~\ref{fig:spin_pumping_benchmark}(e), we get $G_{SP,eff}^{z} = G_{SP}^z G_{SHE}^z/(G_{SP}^z + G_{SHE}^z) = lw (2e^2/h) g_{eff}^{\uparrow \downarrow}$. Hence, we get
\begin{equation}
g_{eff}^{\uparrow \downarrow} = \frac{g^{\uparrow \downarrow}}{1+\frac{\lambda}{\sigma}\,\frac{2e^2}{h} g^{\uparrow \downarrow} coth \left(\frac{t}{\lambda}\right) },
\label{eq:gmix_eff}
\end{equation}
which matches the mathematical expression derived in literature~\cite{RefWorks:876}. The above equation can be backcalculated to get the the bare spin mixing conductance $g^{\uparrow \downarrow}$ with the inequality $(2e^2\lambda/h\sigma) g_{eff}^{\uparrow \downarrow} \, coth (t/\lambda) < 1$, since $g^{\uparrow \downarrow} > 0$. Note that $g_{eff}^{\uparrow \downarrow}$ (and not the bare $g^{\uparrow \downarrow}$) can be determined from the enhancement of damping in ferromagnetic resonance experiments~\cite{RefWorks:814,*RefWorks:813}.

Figure~\ref{fig:spin_pumping_benchmark}(g) shows the charge circuit for the generation of inverse spin Hall voltage~\cite{RefWorks:1253} ($V_{ISHE} = V_2^c - V_1^c$), which also considers the conductance $G_{SH} = \sigma_{mag} t_{mag} w/l$ due to current shunting through the magnet, if it is metallic. The charge current sources in Fig.~\ref{fig:spin_pumping_benchmark}(g), depend on the spin potential difference between the nodes 3 and 4 in the  Fig.~\ref{fig:spin_pumping_benchmark}(c) as $I_0^c = \beta G_0 \left(V_3^z - V_4^z \right)$, where $G_0 = \sigma t w/l$, $\beta = \theta_{sh} l/t$, and $\theta_{sh}$ is the spin Hall angle~\cite{RefWorks:1253}. 

Applying KCL at node 1 of the charge circuit in Fig.~\ref{fig:spin_pumping_benchmark}(g), we get $I_0^c = \left(V_1^c - V_2^c \right) \left(G_0 + G_{SH} \right)$ and hence
\begin{equation}
V_{ISHE} = - \beta \left(\frac{G_0}{G_0 + G_{SH}}\right) \left(V_3^z - V_4^z \right).
\label{eq:V_ISHE_def}
\end{equation}

To calculate $\left(V_3^z - V_4^z \right)$, we apply KCL at nodes 3 and 4 of the spin circuit in Fig.~\ref{fig:spin_pumping_benchmark}(c), and we get
\begin{equation}
V_3^z =  \frac{G_1^z + G_2^z}{D} V_{SP}^{DC} G_{SP}^z, \quad
V_4^z =  \frac{G_2^z}{D} V_{SP}^{DC} G_{SP}^z,
\label{eq:KCL_sol}
\end{equation}
where $D=\left(G_1^z + G_2^z \right) \left(G_{SP}^z + G_1^z + G_2^z \right) - \left(G_2^z\right)^2$. From Equation~\eqref{eq:V_ISHE_def}, we get
\begin{equation}
V_{ISHE} = - \frac{\theta_{SH}l \lambda e \tilde{S} \omega g^{\uparrow \downarrow} sin^2\theta tanh\left(\frac{t}{2\lambda}\right)}{2\pi \left(\sigma t + \sigma_{mag} t_{maag} \right) \left(1+\frac{\lambda}{\sigma}\,\frac{2e^2}{h} g^{\uparrow \downarrow} coth \left(\frac{t}{\lambda}\right)  \right)},
\label{eq:V_ISHE}
\end{equation}
which matches the mathematical expression derived in literature~\cite{RefWorks:814,*RefWorks:813}.

Figure~\ref{fig:spin_pumping_multilayer} shows spin pumping in two subsequent layers and the corresponding spin circuits to determine the effective spin mixing conductance of the whole structure. The spin conductance of the NM layer 2 in the Fig.~\ref{fig:spin_pumping_multilayer}(c) can be determined as $G_{s} = G_1^s + G_1^s G_2^s/(G_1^s + G_2^s) = G_\lambda^s tanh(t^s/\lambda^s)$ where $G_\lambda^s =\sigma^s l w/\lambda^s$ and the whole spin conductance of the two NM layers can be written as $G_{L} = G_1 + G_2 (G_1+G_s)/(G_2 + (G_1+G_s))$. Accordingly, we get
\begin{equation}
\frac{1}{G_{L}} = \frac{1}{G_{\lambda}} \frac{G_\lambda cosh\left(\frac{t}{\lambda}\right) + G_\lambda^s sinh\left(\frac{t}{\lambda}\right)tanh\left(\frac{t^s}{\lambda^s}\right)}{G_\lambda sinh\left(\frac{t}{\lambda}\right) + G_\lambda^s cosh\left(\frac{t}{\lambda}\right)tanh\left(\frac{t^s}{\lambda^s}\right)},
\label{eq:G_L_match}
\end{equation}
where $G_\lambda = \sigma l w/\lambda$ and this expression matches the mathematical expression derived in literature~\cite{RefWorks:1090}. For more complex multilayers, the analytical expression is tedious and this explains the prowess of the spin circuit approach. We can simply write a netlist (see the node numbers in squares in the Fig.~\ref{fig:spin_pumping_multilayer}(b)) as follows for the two layer case.
\begin{lstlisting}[mathescape,columns=fullflexible,basicstyle=\fontfamily{lmvtt}\selectfont,]
	conductances = [1 2 $G_{SP}$;
			2 0 $G_1$; 2 3 $G_2$; 3 0 $G_1$;
			3 0 $G_1^s$; 3 4 $G_2^s$; 4 0 $G_1^s$]
	voltageSources = [ 1 0 $V_{SP}$]
\end{lstlisting}
With a circuit solver, we can solve for $V_3^z$ (also the 3-component circuit in Fig.~\ref{fig:spin_pumping_benchmark}(b) and and the 4-component circuit in Fig.~\ref{fig:spin_pumping} can be solved in general) and using $\left(V_{SP}-V_3^z\right)G_{SP}= V_{SP} G_{SP,eff}$, we can determine the effective spin mixing conductance of the whole structure $G_{SP,eff}$ (shown in Fig.~\ref{fig:spin_pumping_multilayer}(e)) as 
\begin{equation}
G_{SP,eff}=\left(1-\frac{V_3^z}{V_{SP}}\right) G_{SP}.
\label{eq:G_SP_eff}
\end{equation}

To summarize, we have developed the spin circuit representation of spin pumping and have shown that such representation accords to the established mathematical expressions in literature. Such circuits can be simply solved analytically and when more complex they can be solved programmatically to analyze and propose complex devices. We have not considered conductance for the interfacial Rashba-Edelstein effect at NM-NM interface~\cite{RefWorks:1254,*RefWorks:1257,*RefWorks:1258} or any significant interfacial spin memory loss at FM-NM interface, on which (and also on spin diffusion length, spin Hall angle) there is controversy~\cite{RefWorks:1339,*RefWorks:1042,*RefWorks:1316,*RefWorks:986,*RefWorks:1114,*RefWorks:1319,*RefWorks:1113,*RefWorks:1016,*RefWorks:983,*RefWorks:980}. It needs to also carefully consider the low-thickness regime and the effect of magnetic proximity effect~\cite{RefWorks:1005,*RefWorks:1006,*RefWorks:1312,*RefWorks:1007,*RefWorks:988,*RefWorks:1323}. The spin circuit presented here has immense consequence on the development of spintronic technology.

\vspace*{2mm}
This work was supported by FAME, one of six centers of STARnet, a Semiconductor Research Corporation program sponsored by MARCO and DARPA. 

%

\end{document}